\documentclass[11pt,twoside]{article}
\usepackage{asp2010}

\resetcounters

\begin{document}

\title{The Galactic M Dwarf Flare Rate}
\author{Eric~J.~Hilton$^1$, Suzanne~L.~Hawley$^1$, Adam~F.~Kowalski$^1$, and Jon~Holtzman$^2$}
\affil{$^1$University of Washington, Box 351580, UW, Seattle WA, 98195, USA}
\affil{$^2$New Mexico State University, P. O. Box 30001, MSC 4500, Las Cruces NM, 88003, USA}

\begin{abstract}
M dwarfs are known to flare on timescales from minutes to hours, with flux increases of several magnitudes in the blue/near-UV. 
These frequent, powerful events, which are caused by magnetic reconnection, 
will have a strong observational signature in large, time-domain surveys. 
The radiation and particle fluxes from flares may also exert a significant influence on the atmospheres of orbiting planets, 
and affect their habitability. 
We present a statistical model of flaring M dwarfs in the Galaxy that allows us to predict the observed flare rate along 
a given line of sight for a particular survey depth and cadence. 
The parameters that enter the model are the Galactic structure, the distribution of magnetically active and inactive M dwarfs, 
and the flare frequency distribution (FFD) of both populations. 
The FFD is a function of spectral type, activity, and Galactic height. 
Although inactive M dwarfs make up the majority of stars in a magnitude-limited survey, 
the FFD of inactive stars is very poorly constrained. 
We have organized a flare monitoring campaign comprising hundreds of hours of new observations from 
both the ground and space to better constrain flare rates. 
Incorporating the new observations into our model provides more accurate predictions of stellar variability caused by flares on M dwarfs. 
We pay particular attention to the likelihood of flares appearing as optical transients (i.e., host star not seen in quiescent data).
\end{abstract}

\section{Introduction}

Stellar flares are powered by magnetic reconnection events.
Although most last for only a few minutes, they have been observed to last for up to 8 hours \citep{Kowalski2010}.
Their strongest observational signal is a flux increase in the blue/near-UV, which can be several magnitudes for the
biggest flares.
Because M dwarfs account for such a large fraction of the stars in the Galaxy, photometric variability on these
stars will have an significant signature in time-domain surveys such as Pan-STARRS \citep{Kaiser2004}, PTF \citep{Law2009}, and LSST \citep{LSST2009}. 

M dwarfs are an excellent place to look for extrasolar planets because planets have relatively large RV effects and transit depths on low-mass stars.
Many planets, including several super-Earths, have been discovered orbiting M dwarfs \citep[e.g., ][]{Udry2007,Charbonneau2009,Correia2010}. 
However, magnetic activity on M dwarfs can hamper planet searches, since
starspots and flares introduce both photometric noise and radial velocity jitter \citep{Wright2005,Basri2010,L'opez-Santiago2010}.
At the same time, the intense flux of high energy photons and particles caused by flares may have a significant impact on the 
atmospheres of Earth-like planets in the close-in habitable zones of M dwarfs \citep{Buccino2007,Segura2010}.
 
Our group has used the serendipitous observations of flares in surveys to make estimates of the flare rate.
\citet{Kowalski2009} searched for flares in the $\sim$80 photometric epochs from the SDSS Stripe 82 survey \citep{Ivezi'c2007},
while \citet{Hilton2010} identified flares in the sample of M dwarfs with SDSS spectroscopy \citep{West2008}.
These studies confirmed that while the flare rate increases with later spectral subtype, the flare energy on earlier type stars
was higher.
We also found that the flare stars were preferentially closer to the Galactic midplane, which we interpret as an age effect,
with younger stars flaring more frequently.

Although the occurrence of an individual flare cannot be predicted, dedicated photometric monitoring campaigns of individual
stars can yield measurements of the frequency of flares of various energies, also called the flare frequency distribution (FFD).
\citet{Moffett1974} and \citet{Lacy1976} obtained FFDs on several of the most well-known and most prolific flare stars in the Solar neighborhood,
nearly all of which are active mid-type M dwarfs.
We have extended these measurements of the FFD to inactive stars, as well as both early- and late-type M dwarfs.

\section{Observations}
Measuring the FFD of a star requires time-resolved photometric
monitoring over the course of many flares.
We have used four telescopes (described in Table 1) to obtain almost 500 hours of flare monitoring of our targets.

\begin{table}[!ht]
\caption{The telescopes used to collect flare-monitoring data}
\smallskip
\begin{center}
{\small
\begin{tabular}{llll}
\tableline
\noalign{\smallskip}
Location & Size & Operator & Filters \\
\noalign{\smallskip}
\tableline
\noalign{\smallskip}
Apache Point Observatory & 3.5m & Astrophysical Research Consortium & {\it B}  \\
Apache Point Observatory & 1.0m & New Mexico State University & {\it U}  \\
Manastash Ridge Observatory & 0.8m & University of Washington & {\it u,g}  \\
Apache Point Observatory & 0.5m & Astrophysical Research Consortium & {\it u}  \\
\noalign{\smallskip}
\tableline
\end{tabular}
}
\end{center}
\end{table}

We chose bright targets in order to have higher time resolution while maintaining high signal-to-noise.
The cadence of our observations was typically about ten seconds, except for the NSMU 1.0m telescope, 
which has a longer readout time and a cadence of approximately 60 seconds. 
The data reduction and aperture photometry was done with standard IRAF tasks.
Because we are only concerned about variation in the lightcurve, we used differential photometry.
We use the equivalent width of H$\alpha$ to classify each of our targets as either magnetically active or inactive.
Following the criteria of \citet{West2008}, stars with H$\alpha$ equivalent width of more than 1$\AA$ in emission are considered active.

\section{Flare Frequency Distributions}

In order to derive the flare frequency distributions for each bin of spectral type and activity level, 
we first need to identify flares, calculate the flare energies, transform each of these flare energies to a common filter,
and then combine the observations from different nights and different stars.

\subsection{Flare Identification and Energies}

For each night of observations, we first define a period of quiescence.
This is identified by eye and is at least an hour in duration.
We use this quiescent period to define the quiescent flux (taken as the median flux value)
and the precision of our data, taken as the standard deviation of the data points during the quiescent period.
We define a flare as having occurred if there are at least three consecutive measurements more than 2.5 $\sigma$ above the
quiescent value, with at least one of those measurements being $\geq$ 5 $\sigma$ above the quiescent value.
The flare start and stop times are identified as when a running average of several measurements is equal to the quiescent value.
Since this procedure requires the flux to return to the quiescent value before the flare is determined to have ended, 
groups of flares or subpeaks within a flare are typically identified as one flare.  

We follow the method of \citet{Gershberg1972, Moffett1974, Lacy1976} to find the flare energies in the observed filter.
We first calculate the flare flux as 
\begin{equation}
I_{f}(t) = \frac{I_{Flare}(t) + I_0}{I_0} - 1
\end{equation}
where $I_0$ is the quiescent flux.
The equivalent duration, $P$, is simply the time integral between the time of flare start and stop (described above).
\begin{equation}
P = \int \! I_{f}(t) \, dt
\end{equation} 

The total flare energy in filter $x$ is
\begin{equation}
E_x = q_xP
\end{equation}
where $q_x$ is the quiescent flare luminosity in filter $x$.
The quiescent flare luminosity is easily calculated from the known absolute magnitude of the star.
The absolute magnitudes come from the Palomar/MSU Nearby Star Spectroscopic Survey \citep{Reid1996,Hawley1997,Gizis2002,Reid2002},
or the Gleise catalogue \citep{Gliese1969}.
We transform the energies of all flares into SDSS {\it u} using our own simultaneous observations and a relationship from \citet{Lacy1976}.
The total observing time, number of stars, and number of flares in each bin is given in Table 2.

\begin{table}[!ht]
\caption{Flare Monitoring Targets}
\smallskip
\begin{center}
{\small
\begin{tabular}{lrrr}
\tableline
\noalign{\smallskip}
Bin & Total (hrs) & \# of Flares & \# Stars/Bin \\
\noalign{\smallskip}
\tableline
\noalign{\smallskip}
Inactive M0-2 	& 	201.05	& 5	& 9 \\
Inactive M3-5 	&	107.96	& 4	& 3	\\
Active M3-5		& 	123.15	& 43	& 7 \\
Active M6-9		& 	60.27	& 29	& 4 \\
	
\noalign{\smallskip}
\tableline
\noalign{\smallskip}
Total & 492.43 hours & 81 flares & 23 stars \\
\noalign{\smallskip}
\tableline

\end{tabular}
}
\end{center}
\end{table}

\subsection{Constructing the FFD}

The FFD is a cumulative distribution of flare energies such that the FFD evaluated at energy $E$
is the number of flares per unit time with energy greater than $E$.
To construct the FFD, we need not only the energy of each flare, but also the duration of observations during which a 
flare of that energy is able to be detected.
Because conditions change from night to night, and because we must bin together stars of different magnitudes
with different comparison stars, the minimum detectable flare energy is different for each night on each star.

For each bin, the FFD is constructed from the list of flare energies in that bin and the duration ($\tau$) and 
minimum observable flare energy ($E^{min}$)
for each observing period for stars in that bin.
The differential flare frequency is the inverse of the sum of all observing periods whose minimum observable energy is less than the flare energy.

\begin{equation}
FF(E_{i}) = 1/ \sum_{E_{\hspace{2mm}i}^{min}}^{E_{\hspace{2mm}max}^{min}} \tau_{i}
\end{equation}
where $E_{\hspace{2mm}i}^{min}$ is the minimum observable flare energy on a night with duration $\tau_{i}$,
and $E_{\hspace{2mm}max}^{min}$ is the night with the largest threshold for detecting a flare.
The cumulative FFD is simply the cumulative sum of the flare frequencies.

The cumulative FFDs for each of our bins are shown in Figure 1.
Previous studies \citep[e.g., ][]{Lacy1976,Audard2000} have described the FFD as a linear relation between the log of the flare 
energy and the log of the frequency
\begin{equation}
\label{eq:lin_fit}
\log \nu = \alpha + \beta \log E_U
\end{equation}
Overplotted in Figure 1 is our best fit for each FFD.

\begin{figure}[!ht]
\plotone{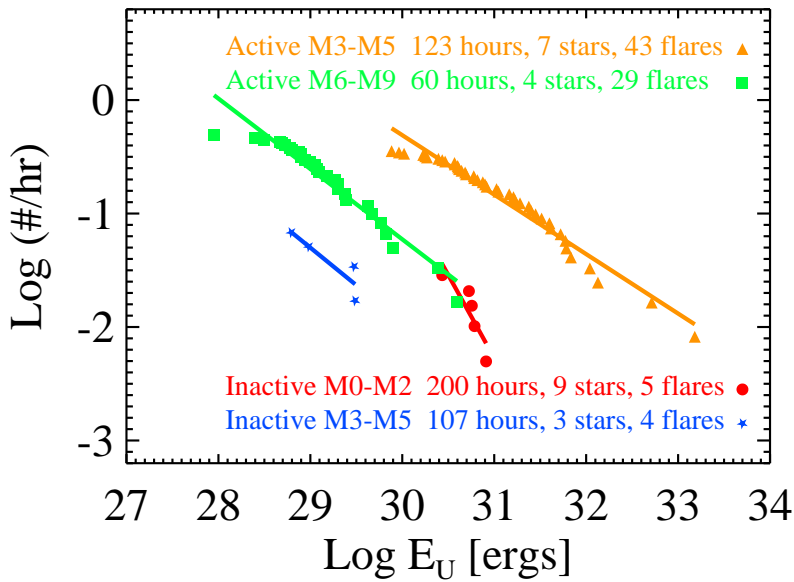}
\label{fig:ffd}
\caption{Using nearly 500 hours of flare monitoring observations, we calculate
 the FFDs of four bins of differing spectral type and activity status are shown here, along with 
the best fit model.
We find that inactive stars flare less frequently than active stars. }
\end{figure}

The FFD for the active M3-M5 stars is consistent with the measured FFDs of several of the most well-known flare stars 
(all active and between M3 and M5) from \citet{Lacy1976}.
The FFDs for the other three bins represent the first measurement for these bins.
We note that inactive stars flare much less frequently than active stars, and with less energy.
For the inactive stars, flares on earlier type stars are more energetic than on later type stars.
This trend is also apparent for the active stars,
if the FFD of the only measured active early M star, Au Mic \citep[][not shown]{Lacy1976}, is representative of all active early M stars.
Although it may be that flares on earlier type stars are intrinsically more energetic, this is at least partially a selection effect,
since a flare on an earlier type star will cause a smaller change in apparent magnitude 
than a flare of equal energy on a later type star, the so called ``contrast effect.''

\section{Galactic Flare Model}

We would like to predict the number and size of flares that will be seen in time domain surveys. 
To do this, we create a model of flaring in the Galaxy.
The first step in modeling the Galactic flare rate is modeling the light curves of individual stars.
We adopt the classic flare shape \citep[see e.g.,][]{Moffett1974,Hawley1991, Kowalski2010} of an impulsive phase 
with a linear flux rise and partial linear decrease, followed by a gradual phase consisting of an exponential flux decay.
We insert flares into a light curve by drawing from the measured FFD for a particular type of star.
Using the model light curves, we calculate the fraction of time a star is seen at increased brightness.
The black line in Figure 2 (top panel) shows the FFD for the active M3-M5 bin, while the black line in
the bottom panel shows the fraction of time a hypothetical dM3.5e star is seen at increased brightness from flares.
The uncertainty in the measurement of the FFD (shaded gray) is dominated by our choice of the flare definition, 
especially the start and stop times, 
and the assumption that all stars within a particular bin have the same intrinsic FFD.
Different input FFDs result in significant changes in the amount of time spent at increased brightness. 
In principle, we can use time domain surveys, which provide measurements of the fraction of time spent at increased brightness, 
to determine which FFD is the best fit to the survey data.

\begin{figure}[!ht]
\plotone{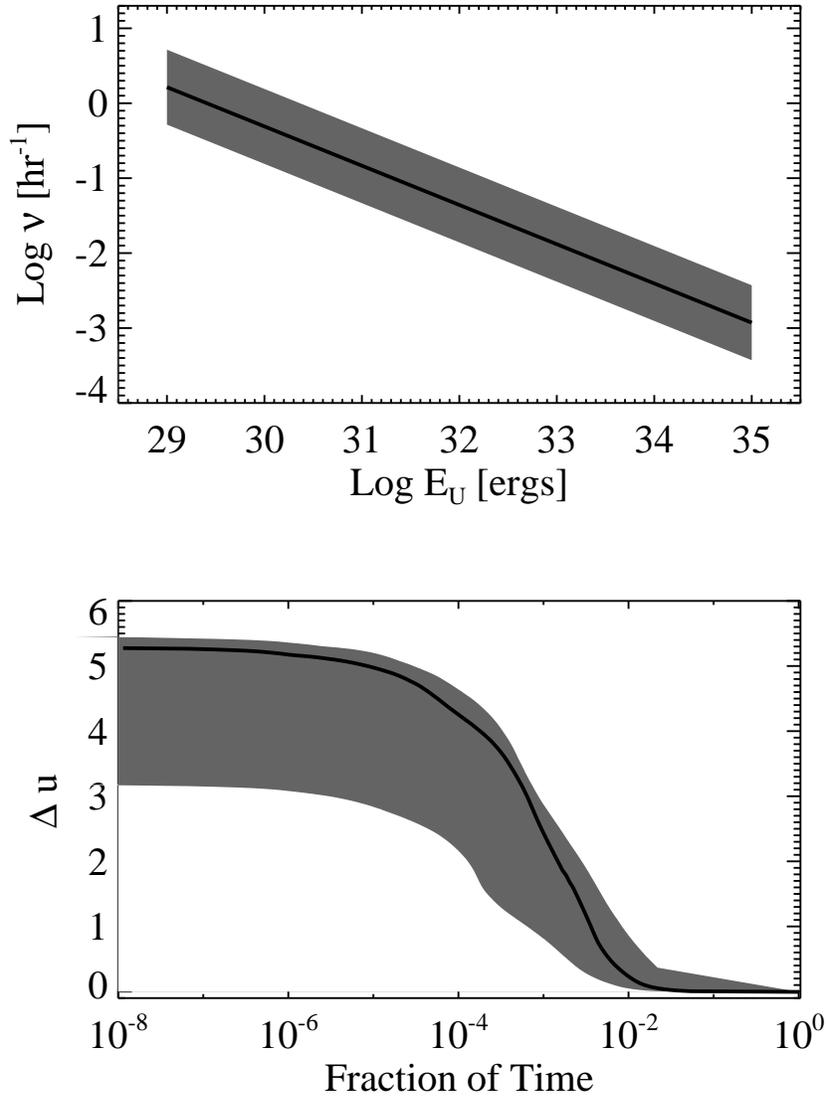}
\caption{Top: The measured FFD of an active M3.5, with the uncertainty shown as shaded gray.
Bottom: The fraction of time an active M3.5 spends at increased brightness.
Stars with different FFDs spend vastly different amounts of time at highly enhanced magnitudes.
Future studies may be able to use the large number of flare observations from time domain surveys 
to determine the FFD more precisely.
}
\end{figure}

We have created a model of flaring on M dwarfs in the Galaxy to predict the number and magnitude of flares that will be observed in time domain
surveys, with a particular focus on LSST.
We generate a model Galaxy with thin disk, thick disk, and halo components \citep{Juric2010}.
The scale heights, lengths, and densities of these components have been determined from SDSS star counts \citep{Juri'c2008,Bochanski2010}.
The active fractions of each spectral subtype as a function of Galactic height are extrapolated from \citet{West2008}.

For a given pointing and field of view size, the Galactic model determines the number, color, distance, and activity level of M dwarfs along that line of sight.
The activity status and spectral type determine the FFD used to generate a light curve for each star in the model.

To verify the utility of this model, we compared our model results to the measured flare rates \citep{Kowalski2009} from the 
SDSS Stripe 82 photometric catalogue \citep{Ivezi'c2007}. 
Each of the $\sim$50,000 M dwarfs in Stripe 82 was observed $\sim$80 times, with 271 epochs meeting stringent flare criteria \citep{Kowalski2009}.
The magnitude distribution of these flares is shown in Figure 3 (red line).
Adopting the same number of epochs, field of view, flare criteria, and survey depth of the Stripe 82 data in our flare model, we predict 283 flares,
which are overplotted in Figure 3 (blue line).
Our model not only matches the total number of flares, but independently reproduces the distribution of flare magnitudes remarkably well.

\begin{figure}[!ht]
\plotone{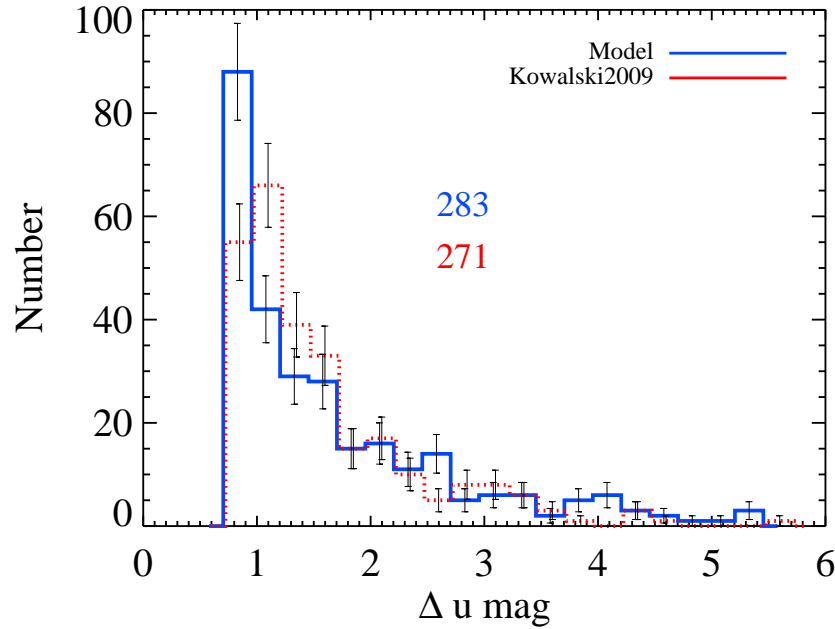}
\caption{The magnitude distribution of flares measured in SDSS Stripe 82 from \citet{Kowalski2009} (red line) and
the model prediction using the same survey parameters (blue line). }
\end{figure}

\subsection{Predicting Flare Rates in Time Domain Surveys}

The good agreement between our flare model and the measured Stripe 82 flare rates gives us 
confidence that we can predict the frequency of flares in future time domain surveys.
However, there are three areas where limited data may affect the model predictions.
These limitations are especially important to the prediction of transients --- flares that are bright enough to 
make a star briefly visible that is otherwise too faint to be seen in the survey.

The first uncertainty is the active fraction of stars as a function of spectral type and distance from the Galactic midplane.
The \citet{West2008} results are based on SDSS spectroscopic data, which are limited to bright objects.
The active fraction at late spectral types and at far distances, which will certainly be visible to LSST, are currently
unconstrained, and will likely remain so in the near-term. 
The second reason is that a study of flares in the SDSS spectroscopic survey finds that flare stars are preferentially
closer to the Galactic plane than the active stars \citep{Hilton2010}, indicating that the FFD of active stars may be a function of distance
from the Galactic plane. 
The third limitation to our model inputs is in the rate of flares at the high energy, low frequency end of the distribution. 
Because large flares happen infrequently, we have limited information on the frequency of the highest energy flares.
We don't know if there is a break in the distribution, or simply a cut-off.

We have used the expected survey parameters for LSST to predict the number of flares seen in a single 15-second
{\it u}-band exposure.
Figure 4 shows the number of flares seen as a function of Galactic latitude. 
The decrease with increasing latitude is primarily caused by the smaller number of stars in sightlines looking out of the Galactic plane.
As expected, there are more small flares than large flares.
The error bars are calculated using the uncertainties in the FFDs (see Figure 2).
The number of flares seen in a single night of LSST observing will be larger than the entirety of the \citet{Kowalski2009} SDSS Stripe 82 results.
In the future, we can determine the FFD of stars to a much greater precision,
and for much smaller spectral type and Galactic height bins, by varying our input FFDs in the model to fit the large number of LSST flare observations. 

\begin{figure}[!ht]
\plotfiddle{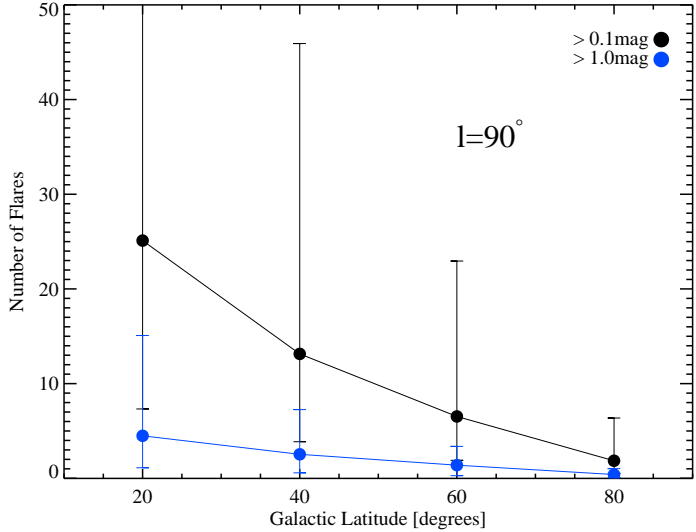}{2.6in}{0.}{50.}{50.}{-150}{0}
\caption{The predicted number of flares that will be seen in a single LSST {\it u}-band exposure.
The decrease with Galactic latitude is caused by a decrease in the number of M dwarfs in the sightline.}
\end{figure}

Astronomers interested in exotic transients can neglect most stellar flares, since the source of the flare is known from the quiescent photometry.
However, it is certainly possible for an M dwarf to have such a large flare that it will appear as an optical transient.
In this case, the star is too faint to be visible in quiescence in any filter, but is briefly bright enough that it is seen during the flare, and will 
thus appear as a blue point source with no observed host.
To meet this criteria, a flare must have a large change in magnitude:
\begin{eqnarray}
\Delta u > (u_q - z_q) + (z_q - z_{lim}) + (z_{lim} - u_{lim}) = u_q - u_{lim} \\ 
\mbox{where } z_q < z_{lim}
\end{eqnarray}
where the subscript $q$ represents quiescent measurements and the subscript $lim$ represents the survey limiting
magnitude in that filter.
The first term is the $u-z$ color of the star, which is several magnitudes for late M dwarfs.
The second term means that stars close to the survey limit more likely to appear as transients, 
and the final term is the difference between the survey limits in the two filters.
Note that although we use {\it u} and {\it z} filters in this example, we can make similar predictions for any other filter.

Using the LSST survey criteria, we predict that the probability of M dwarf flares appearing as optical transients in
a single {\it u}-band exposure is $\sim$10\% at low Galactic latitude, decreasing to just a few percent at high latitudes.
Only the largest flares will meet the transient criteria, and as mentioned above, the frequency of very large flares is not well measured.
The error bars are therefore quite large.

\section{Conclusions}

We have collected nearly 500 hours of flare monitoring observations to make the first measurements 
of the FFD of inactive early- and mid-type M dwarfs and for active late-type M dwarfs.
The FFD is best represented as a linear relation between the log of the quiescent flare energy and the 
log of the frequency.
We find that flares on earlier type stars are more energetic than on later type stars, and that
inactive stars flare less frequently than active stars.

We incorporate the measured FFDs into our Galactic M dwarf flare model, which can be used to 
predict the number and magnitude of flares that will be seen in any survey.
To make this prediction, we also adopt a flare shape, an active fraction (which is a function of spectral type
and distance from the Galactic plane), a luminosity function, and a Galactic model.
We draw from the appropriate FFD to generate a light curve for each star along a given sightline. 
Using the survey parameters from SDSS Stripe 82, our model was able to reproduce both the number
and the magnitude distribution of flares found by \citet{Kowalski2009}. 

Applying our model to the survey parameters of LSST predicts tens of flares $>0.1$ magnitudes and several flares $>1.0$ magnitudes
in each {\it u}-band exposure.
We can also predict the probability of observing a large flare occurring on a star that is too faint to be seen during quiescence.
These large flares, which appear as optical transients in the survey, have a few percent probability of being observed
in each {\it u}-band exposure, although the uncertainties are large.

Finally, when large numbers of flares are measured in time domain surveys such as Pan-STARRS, PTF, and LSST, 
our model can be used to precisely determine the FFD of stars in very small bins of spectral type or color and distance from the Galactic plane.
This will inform our knowledge of how the magnetic field of M dwarfs changes with spectral type (in particular how it changes across the 
fully convective boundary), with age, and with activity level.

\bibliography{hilton_e}

\end{document}